\documentclass[preprint2]{aastex}
\def\stacksymbols #1#2#3#4{\def\theguybelow{#2}
	\def\verticalposition{\lower#3pt}
	\def\spacingwithinsymbol{\baselineskip0pt\lineskip#4pt}
	\mathrel{\mathpalette\intermediary#1}}
\def\intermediary #1#2{\verticalposition\vbox{\spacingwithinsymbol
	\everycr={}\tabskip0pt
	\halign{$\mathsurround0pt#1\hfil##\hfil$\crcr#2\crcr
		\theguybelow\crcr}}}
\def\lta{\stacksymbols{<}{\sim}{2.5}{.2}}
\def\gta{\stacksymbols{>}{\sim}{3}{.5}}

\begin{document}

\title{{\it Chandra} DETECTION OF MASSIVE BLACK HOLES IN 
GALACTIC COOLING FLOWS}


\author{Fabrizio Brighenti$^{1,2}$ and William G. Mathews$^1$}

\affil{$^1$University of California Observatories/Lick Observatory,
Board of Studies in Astronomy and Astrophysics,
University of California, Santa Cruz, CA 95064\\
mathews@lick.ucsc.edu}

\affil{$^2$Dipartimento di Astronomia,
Universit\`a di Bologna,
via Ranzani 1,
Bologna 40127, Italy\\
brighenti@bo.astro.it}






\vskip .2in


\begin{abstract}

Anticipating forthcoming {\it Chandra} X-ray observations, 
we describe the continuation of interstellar
cooling flows deep into the cores of elliptical galaxies. 
Interstellar gas heated to $T > 1$ keV in the potential 
of massive black holes ($r \lta 50$ pc) should be visible unless 
thermal heating is diluted by non-thermal pressure.
Since our flows are subsonic near the massive black holes, 
distributed cooling continues within $\sim 300$ pc.
Dark, low mass stars formed in this region may be responsible 
for some of the mass attributed to central black holes.

\end{abstract}


\section{INTRODUCTION}

Considerable observational evidence now exists 
for supermassive black holes (SMBHs) in the 
centers of giant elliptical galaxies.
These mass concentrations, long suspected to be the origin of 
non-thermal energy in active galaxies, 
are now thought to be 
responsible for large gas or stellar velocities 
in galactic cores (Ho 1998).
We discuss here the interaction of SMBHs with  
hot ($T \sim 10^7$ K) interstellar gas.
The local compression and heating of interstellar  
gas in the SMBH potential can be detected with the 
{\it Chandra} X-ray telescope.
Indeed, the inflow of interstellar gas 
may explain the growth of SMBHs in galactic cores.
However, X-ray detection of SMBHs may not 
be possible if 
the non-thermal interstellar pressure 
is comparable to the pressure of hot thermal gas.

As we were completing this work, a similar study of the 
SMBH-gas interaction by Quataert \& Narayan (1999) (QN) 
appeared on the astro-ph preprint server. 
While some of the conclusions of QN are supported 
by our flow calculations, 
our results are qualitatively different 
in a number of important ways.
The steady state flows considered by QN are constrained 
to pass through a sonic point, 
becoming supersonic near the SMBH.
By comparison, 
our (non-steady) computational gas flows near the SMBH 
adapt naturally to the interstellar  
cooling flow in the surrounding galaxy and, as a result, 
our flow velocity is fully subsonic near the SMBH. 
We also consider several additional physical effects 
not discussed by QN: time-dependent 
ejection of mass by evolving stars, 
heating by Type Ia supernovae, non-thermal pressure, 
and the gravitational field of 
(dark) cooled interstellar gas which 
in some observations may be confused with the mass of the SMBH.

Although we use observations of the bright 
Virgo elliptical NGC 4472 as a guide for our models, we 
do not seek here a set of parameters for an 
optimal fit to this galaxy. 
Instead, our objective is to illustrate some of the physical effects 
that must be considered in interpreting X-ray observations 
of the cores of bright ellipticals with the {\it Chandra} telescope.
At a distance $d = 17$ Mpc appropriate for NGC 4472, 
{\it Chandra's} 
resolution ($\sim 0.5$ arcseconds) is $\sim 40$ pc.

\section{CONSTRUCTING MODELS}

The computational procedure we follow here is identical to 
that described in our recent papers (e.g. Brighenti \& Mathews 1999a).
With NGC 4472 in mind, we construct a galaxy in a cosmological 
environment beginning with an overdensity perturbation 
designed to form a galaxy group. 
Stars of total mass $M_{*t} = 7.26 \times 10^{11}$ 
$M_{\odot}$ (appropriate to NGC 4472  
with $M/L_B = 9.2$; van der Marel 1991),
are formed from the baryons at an early time 
($t_{*s} = 1$ Gyr) and later 
($t_* = 2$ Gyrs) merge into a de Vaucouleurs profile. 
An evolving NFW dark halo develops around the galaxy over time,
accompanied by secondary infall, although 
within an effective radius ($r_e = 8.57$ kpc for NGC 4472) 
stellar mass loss provides about half 
of the hot interstellar gas.
A cooling flow develops in the hot interstellar gas 
at early times and slowly evolves toward its present configuration 
at $t_n = 13$ Gyrs.
Over time a large mass of interstellar gas cools in NGC 4472, 
$M_{do} \approx 
t_n L_x/(5 k T / 2\mu m_p) \approx 
4 \times 10^{10}$ $M_{\odot}$. 
Since this mass exceeds the mass observed in the central 
black hole in NGC 4472, $\sim 2.6 \times 10^9$ $M_{\odot}$
(Magorrian et al. 1998),
interstellar  
gas must cool in localized thermal instabilities distributed 
throughout an extended galactic volume within $r_e$.
This distributed interstellar cooling is modeled by 
adding a sink term to the equation of continuity:
$-q(r) \rho /t_{do} \propto \rho^2,$
where $t_{do} = 5 m_p k T / 2 \mu \rho \Lambda$
is the time for gas to cool locally by radiative 
losses at constant pressure (e.g. Sarazin \& Ashe 1989).
We find that taking $q = 1$ provides a satisfactory 
global interstellar flow solution for NGC 4472 
that matches currently observed density,
temperature and metallicity profiles reasonably well
(Brighenti \& Mathews 1999a; 1999b);
we adopt $q = 1$ here.

To resolve the small scale flow near the SMBH we have 
refined our logarithmic spatial grid in the galactic 
core to an innermost zone of radius 12 pc. 
Interstellar gas is influenced by the black hole within a radius
$r_h \sim G M_{bh}/ c_s^2 \approx 48 (M/10^{9} M_{\odot})
(c_s^2 / 300~{\rm km/s})^{-2}$ pc; $c_s$ is the 
sound speed in the hot gas.
We ignore the effects of galactic rotation.
Although the mass-losing stellar systems in large ellipticals 
like NGC 4472 are slowly rotating, 
their X-ray images do not indicate the flattening expected 
if the hot interstellar gas shared this rotation
(Brighenti \& Mathews 1996).
Evidently, 
efficient cooling dropout or turbulent viscosity 
removes angular momentum from the hot gas.

\section{A ``STANDARD'' MODEL}

Figure 1 shows the 
interstellar density, temperature 
and (ROSAT band) X-ray surface brightness 
evolved to the present time, $t_n = 13$ Gyrs.
During the evolution, new gas and energy are supplied from 
stellar sources: 
(1) the specific stellar mass loss rate is 
$\alpha_*(t) = 4.7 \times 10^{-20} 
[t/(t_n - t_{*s})]^{-1.3}$ sec$^{-1}$ 
and 
(2) gas is heated by Type Ia supernovae at a rate 
SNu$(t) = $SNu$(t_n)(t_n/t)$, 
where SNu$(t_n) = 0.03$ SNIa 
per 100 yrs per $10^{10}$ $L_{B\odot}$ 
(Cappellaro et al. 1997). 
The interstellar gas temperature is nearly virial, 
$T \approx 10^7$ K. 
The stellar density has a de Vaucouleurs profile 
but flattens to $\rho_* \propto r^{-0.9}$ 
within a ``break'' radius $r_b = 206$ pc  
(Faber et al. 1997).

Dashed lines in Figure 1 refer to solutions without 
central black hole and the solid lines refer to solutions
with a central black hole of mass $2.0 \times 10^9$ 
$M_{\odot}$ introduced at $t_* = 2$ Gyrs. 
In this $q = 1$ solution, an additional $0.7 \times 10^9$ 
$M_{\odot}$ flows inside $r = 12$ pc by $t_n = 13$ Gyrs, 
increasing the central dark 
mass to $2.7 \times 10^9$ $M_{\odot}$. 
The gas density (with SMBH) varies approximately 
as $n \propto r^{-5/4}$ over most of the flow. 
For $r \lta 300$ pc, the computed density 
appears to exceed the innermost observations;
we discuss this apparent discrepancy in more detail below.
Light curves in Figure 1 
for $T(r)$ refer to the background flow; heavy curves 
show the projected temperature including  
emission from distributed cooling regions and 
correspond to the temperature observed.
For constant $q$, cooling dropout is important 
throughout the flow and extends to the 
region adjacent to the SMBH. 
The gas temperature rises dramatically within 
about 50 pc from the SMBH -- detection of this 
thermal peak by {\it Chandra} would provide 
additional evidence for a SMBH. 
This temperature spike should be visible 
even when viewed with {\it Chandra} 
in projection against the entire cooling flow: 
the mean projected thermal temperatures  
within apertures of $R = 50$, 100 and 500 pc 
are 2.95, 2.70 and 2.00 $\times 10^7$ K (with SMBH) 
and 1.04, 1.05 and 1.09 $\times 10^7$ K (without SMBH) 
respectively.

The upper panel in Figure 2 shows the 
current flow velocity $v(r)$ and 
the local mass flow ${\dot M}(r) = 4 \pi r^2 \rho v$. 
A small influence due to the SMBH potential at times $t \sim t_*$ 
can be seen in the distant flow $r \gta 30$ kpc where 
the flow time $t_f = r/v \gta t_n - t_*$.
Overall, the flow is very subsonic 
even in the vicinity of the SMBH where 
$c_s \gta 300$ km s$^{-1}$.
This occurs for two reasons: (1) the high pressure 
in the central thermal spike (e.g. Holzer \& Axford 1970), 
enhanced by the SMBH, and 
(2) mass dropout which tends to make subsonic flow 
even slower.
By contrast, the solutions of QN 
(with $\alpha_*(t) = 0$) 
are constrained to pass through a sonic transition, 
causing the gas to accelerate 
through the thermal spike. 
[Without mass dropout ($q = 0$) we also find a 
transition to supersonic flow at $\sim 60$ pc.] 
Our calculations shown in Figures 1 and 2 were done 
with boundary condition $u(r = 0) = 0$;
accumulated gas in the central zone cools there. 
To encourage the onset of supersonic flow near the SMBH, 
we repeated the calculation 
allowing the gas to flow freely through a small radius 
(as small as $r = 1$ pc with a very high resolution grid) 
and found that the flow was essentially unchanged from 
that shown in the Figures. 
The fully subsonic character of our solutions 
is therefore quite robust and independent of 
our boundary condition at $r = 0$.
[In principle our flow should become supersonic very near 
the Schwarzschild radius ($r_g \approx 10^{-4}$ pc), 
but this cannot be resolved by our spatial grid and 
it is likely that additional physical effects become 
important as $r$ approaches $r_g$.]

Because our flow in the thermal spike near the SMBH 
is subsonic, the cooling time $t_{do}$ remains comparable to 
the flow time $t_{flow} = r/v$ throughout this region 
and cooling dropout continues unabated at small radii, 
unlike the QN solutions. 
This concentrated mass dropout has 
two important astrophysical consequences: 
(1) the mass deposited by interstellar cooling near 
the SMBH may be comparable to that of the SMBH itself,
and
(2) the rate that mass flows into the SMBH at the origin 
is significantly reduced (by about $\sim 10$) 
below the QN solutions.

The mass of cooled interstellar gas in our model
with SMBH in Figure 2 is large: 
$M_{do}(r) = 2.3 \times 10^9$, $3.5 \times 10^9$ 
and $8.0 \times 10^9$ $M_{\odot}$ within
$r = 50$, 100 and 300 parsecs respectively.
These masses are comparable to SMBH 
masses found by Magorrian et al. (1998) 
and they occur within the small region
($\lta 4$ arcseconds or 330 pc for NGC 4472) where
stellar velocities are enhanced by the SMBH.
The gravitational acceleration due to the SMBH,
the dropout mass, old stars and dark halo are also shown 
in Figure 2; note that the dropout mass dominates
gravitational forces at $r \sim 200$ pc.
In view of the possibility of contamination with 
dropout mass, the true mass of the central SMBH 
may be lower than those quoted by Magorrian et al.
These considerations depend on our assumption 
that interstellar mass dropout continues to the smallest 
galactic radii; at the present time there is no reason 
to question this assumption.

In Figure 3 we illustrate a solution identical to 
that in Figure 2 but with an initial SMBH mass of 
only $M_{bh} = 1 \times 10^9$ $M_{\odot}$; the 
visibility of the thermal spike is noticeably reduced.

\section{``NON-STANDARD'' MODELS}

We now return to the apparent disagreement in Figure 1 between 
our models and the 
innermost observed $\Sigma_x(R)$ and deprojected electron density.

Can the apparent flattening of 
observed $\Sigma_x(R)$ at small radii 
be due to absorption of 
X-rays by cold gas in $r \lta 300$ pc?
The HII gas that produces 
optical emission lines in NGC 4472 
(Macchetto et al. 1996) cannot 
account for the X-ray absorption since 
H$\beta$ emission from such HII clouds in 
pressure equilibrium would be $\gta 1000$ times 
more intense than $L_{H\beta}$ 
observed in this region of NGC 4472.
Suppose instead that 
the absorbing gas is cold HI or H$_2$ and  
is distributed in a uniform disk extending to $r \sim 300$ pc 
oriented perpendicular to the line of sight.
If the disk column density is  
$N \approx 10^{21}$ cm$^{-2}$ 
(i.e. $\tau (1~{\rm keV}) \approx 1$),  
the total disk mass, $\sim 2 \times 10^6$ $M_{\odot}$, 
would be consistent with upper limits on 
observed HI and H$_2$ in NGC 4472,
$M_{cold} \lta 10^7$ $M_{\odot}$ 
(Bregman, Roberts \& Giovanelli 1988;
Braine, Henkel \& Wiklind 1997).
However, in pressure equilibrium with the hot cooling flow gas, 
the density of such cold gas (at $T \sim 15$ K 
Mathews \& Brighenti 1999) is very high, 
$n_{cold} \sim 6 \times 10^5$ cm$^{-3}$, and the 
disk thickness is incredibly small, $h \approx N / n_{cold} 
\sim 6 \times 10^{-4}$ pc. 
Whether such a peculiar and fragile disk could exist 
in NGC 4472 (and other Virgo ellipticals with  
similar cores; 
Brighenti \& Mathews 1997) seems unlikely.
However, optical line emission from the surface 
of the disk ionized by stellar UV
(Mathews \& Brighenti 1999) would exceed observed limits. 
For HII gas at $T \approx 10^4$ K the pressure 
equilibrium density at $r_d = 300$ pc is $n_{HII} \approx 
850$ cm$^{-3}$ and the Stromgren radius is 
$r_s \approx 10^{16}$ cm. 
The total H$\beta$ energy emitted by the disk,  
$L_{H\beta,d} \approx 2 \pi r_d^2 r_s n_{HII}^2 \epsilon_{H\beta} 
\approx 4 \times 10^{39}$ ergs s$^{-1}$, 
is $\sim 6$ times larger than
the total H$\beta$ luminosity from 
NGC 4472, $L_{H\beta} \approx 7 \times 10^{38}$ ergs s$^{-1}$.
This inconsistency cannot be removed with other geometrical  
arrangements of the absorbing cold gas within 300 pc, 
so the apparent inner flattening in Figure 1 is 
probably not due to absorption.

We have also explored the possibility that the
$\Sigma_x(R)$ observations 
shown in Figure 1 were centrally flattened by 
the 4 arcsecond
half power Gaussian PRF (Giaconni et al. 1979) of the 
{\it Einstein} HRI.
However, such a convolution of our computed $\Sigma_x(R)$ 
cannot duplicate the 
core observed in NGC 4472. 
X-ray cores in NGC 4472, NGC 4636 and NGC 4649
(Trincheri, Fabbiano \& Canizares 1986) 
are not artifacts of limited observational
resolution. 

These difficulties led us to explore two additional 
``non-standard'' models for NGC 4472: 
flows (1) with extreme mass dropout inside 300 pc 
or (2) with additional non-thermal pressure in this 
same region.

It is often claimed that mass dropout reduces the 
X-ray surface brightness in the central regions of 
cooling flows, but this is not true in general.
For flows in which most of the gas in the central 
region has advected inward from large radii, 
cooling dropout at large radii 
can reduce the amount of gas and associated 
emission $\Sigma_x(R)$ in the central core.
However, this does not apply to galactic cooling flows 
in which much of the gas in the central flow originates 
from stars within or near this region.
In addition, the X-ray emissivity and $\Sigma_x(R)$ are 
locally enhanced at (denser) cooling sites where
mass dropout occurs. 
To illustrate the futility of lowering the central 
$\Sigma_x(R)$ in NGC 4472 with increased local cooling, 
we show in Figure 3 density and temperature profiles 
for the standard flow (at $t_n$ and with SMBH) but 
with a cooling dropout that truncates the 
background flow density: $q(r) = e^{n(r)/n_m}$ with  
$n_m = 0.006 $ cm$^{-3}$. 
The thermal spike around the SMBH is still present in this 
solution, but the apparent 
$n(r)$ (and $\Sigma_x(R)$) within several 
100 pc is only slightly reduced.
The mean projected temperature within 100 pc 
is $T(R < 100) = 1.9 \times 10^7$ K.

Finally, we consider cooling flow models in which 
$n(r)$ and $\Sigma_x(R)$ in the inner flow are 
lowered by an additional non-thermal pressure 
due to relativistic particles or magnetic fields,  
as suggested by radio emission observed in most 
elliptical cores.
To explore this possibility we show in Figure 3 the standard 
flow (at $t_n = 13$ Gyrs with SMBH) in which an additional pressure 
is introduced in the galactic core:
$P_{tot} = P(1 + 5 e^{-r/2~{\rm kpc}})$, 
where $P$ is the thermal gas pressure.
The density (and $\Sigma_x$) in the cooling flow within 
$\sim 300$ pc 
are reduced by this additional pressure 
and parameters for $P_{tot}(r)$ could be adjusted 
in an {\it ad hoc} fashion to achieve a more perfect fit 
with the observations.
However, the apparent temperature of the thermal 
gas is dramatically 
lowered since the external flow 
is now supported mostly by non-thermal pressure.
The projected apparent temperature within $R = 100$ pc 
is only $T = 0.40 \times 10^7$ K, assuming 
this gas is not heated by collective interactions 
with the relativistic gas. 
The discernibility of the thermal spike near the SMBH is 
now greatly reduced.

\section{FINAL REMARKS AND CONCLUSIONS}

In this study we have examined central cooling flows 
in a luminous elliptical similar to NGC 4472.
We have emphasized the evolution of quiescent 
flows that arise naturally from 
gas and energy supplied by galactic stars 
and by secondary infall in the outer regions. 
We have not explored possible transient interactions  
between the cooling flow and an intermittently active 
galactic core (e.g. Ciotti \& Ostriker 1997).
Our objective here has been to anticipate some of the 
interpretive issues that 
may arise when {\it Chandra} observations become available. 

Our main conclusions are as follows:

\noindent
(1) Some dark matter currently attributed to the central
supermassive black hole (SMBH) may be due to 
a population of non-luminous stars created from cooled 
interstellar gas having a bottom-heavy IMF.
In the standard cooling flow for NGC 4472 with SMBH,   
neutral clouds become 
gravitationally unstable (and truncate the IMF) at masses 
$m_u = 0.2$, 0.3 and 0.8 $M_{\odot}$ 
at $r = 50$, 100 and 300 pc respectively 
(Mathews \& Brighenti 1999).
Such dark stars, possibly having approximately radial orbits,
can also produce a radial variation 
in the global baryonic $M/L$.

\noindent
(2) A successful observation of a high temperature peak
in $r \lta 50$ pc by {\it Chandra} in a nearby bright 
elliptical would provide additional strong evidence for 
the existence of SMBHs. 
However, if additional non-thermal pressure is present in 
elliptical cores the SMBH thermal peak may no longer 
be discernible. 

\noindent
(3) Our central flows ($r \lta 300$ pc) differ from the 
Bondi-type transonic flow described by 
Quataert \& Narayan (1999) since we do not insist on 
a sonic transition. 
The mass flow rate into the SMBH is considerably 
reduced in our models.

\noindent
(4) As long as our standard cooling flow assumptions hold,
we expect the gas density to rise steadily toward the 
SMBH approximately as $n \propto r^{-5/4}$.  
But current observations suggest that such a gas density 
peak is not present.
About half of the gas within 300 pc in NGC 4472 
is supplied by mass loss from stars in this region.
In our standard cooling flow model we have assumed that 
all stellar ejecta rapidly enters the hot interstellar phase 
(in $\lta 10^4$ yrs). 
It has occasionally 
been suggested (e.g. Thomas 1986) that stellar ejecta 
cools before entering the hot gas, effectively reducing 
$\alpha_*$. 
However, we find that the central peak in $n(r)$ and 
$\Sigma_x(R)$ persists even if the stellar mass loss 
rate $\alpha_*$ is reduced at small $r$.
Moreover, dynamical arguments 
and gas temperature and metallicity gradients observed at 
larger radii ($r \lta 3r_e$) strongly indicate that 
cooler, enriched stellar ejecta is conductively melting 
into the hot phase.

\acknowledgments

Studies of the evolution of hot gas in elliptical galaxies 
at UC Santa Cruz are supported by
NASA grant NAG 5-3060 and NSF grant 
AST-9802994 for which we are very grateful. 
FB is supported in part by Grant MURST-Cofin 98.









\vskip.1in
\figcaption[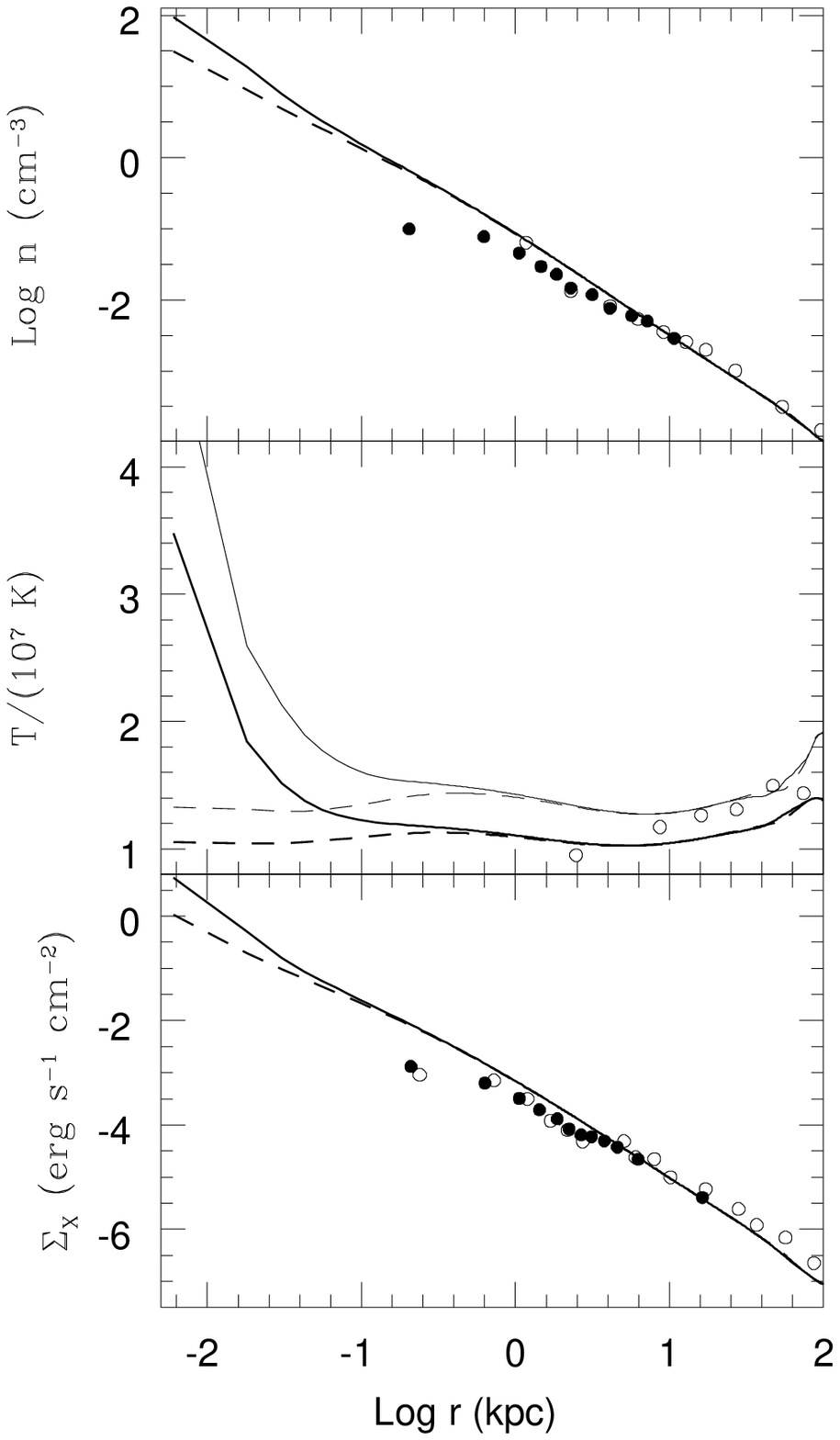]{
Structure of the standard cooling flow solution, 
shown both with ({\it solid line}) and 
without ({\it dashed line}) a supermassive black hole (SMBH) 
of mass $M_{bh} = 2 \times 10^{9}$ $M_{\odot}$.
{\it Top Panel:} Radial variation of apparent gas density 
including cooling regions.
{\it Middle Panel:} Variation of gas temperature with 
galactic radius in the background flow ({\it light lines}) 
and with projected radius including emission from 
cooling regions ({\it heavy lines}).
{\it Bottom Panel:} Variation of ROSAT band X-ray 
surface brightness with projected radius.
Filled circles are {\it Einstein} HRI observations 
(Trinchieri, Fabbiano \& Canizares 1986) and open circles 
are ROSAT HRI and PSPC observations
(Irwin \& Sarazin 1996).
\label{fig1}}

\vskip.1in
\figcaption[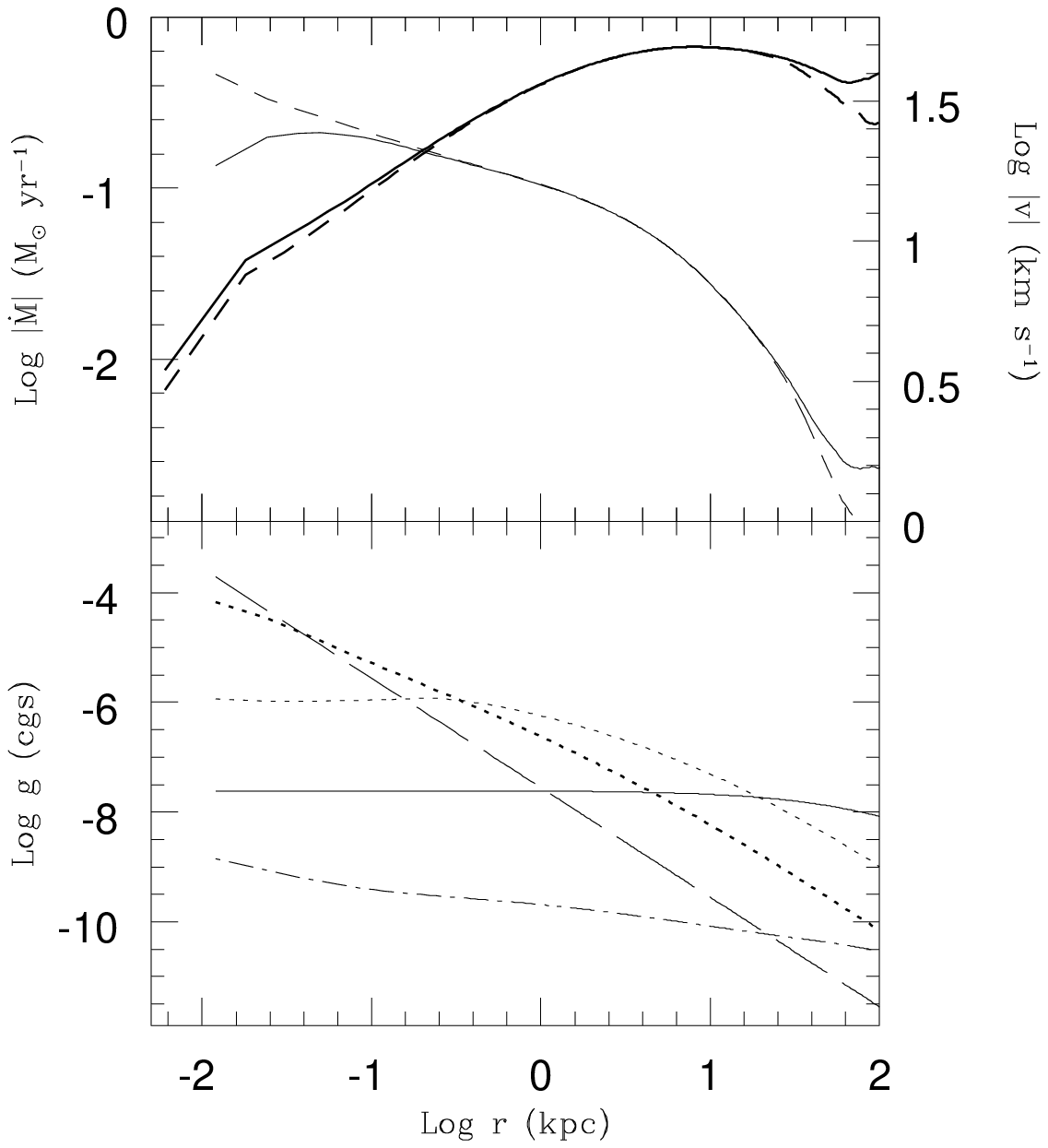]{
Additional properties of the standard cooling flow 
solution shown in Fig. 1.
{\it Top Panel:} Radial variation of the flow velocity 
({\it light lines}) and mass flow rate
({\it heavy lines}).
Solutions with  
and without supermassive black hole (SMBH) are shown 
as solid and dashed curves respectively.
{\it Bottom Panel:} Gravitational acceleration due to 
SMBH ({\it long dashed line}), 
old stellar population ({\it light dotted line}),
stars formed from cooled interstellar gas 
({\it heavy dotted line}),
the dark halo ({\it solid line}) and 
the interstellar gas ({\it dot-dashed line}).
\label{fig2}}

\vskip.1in
\figcaption[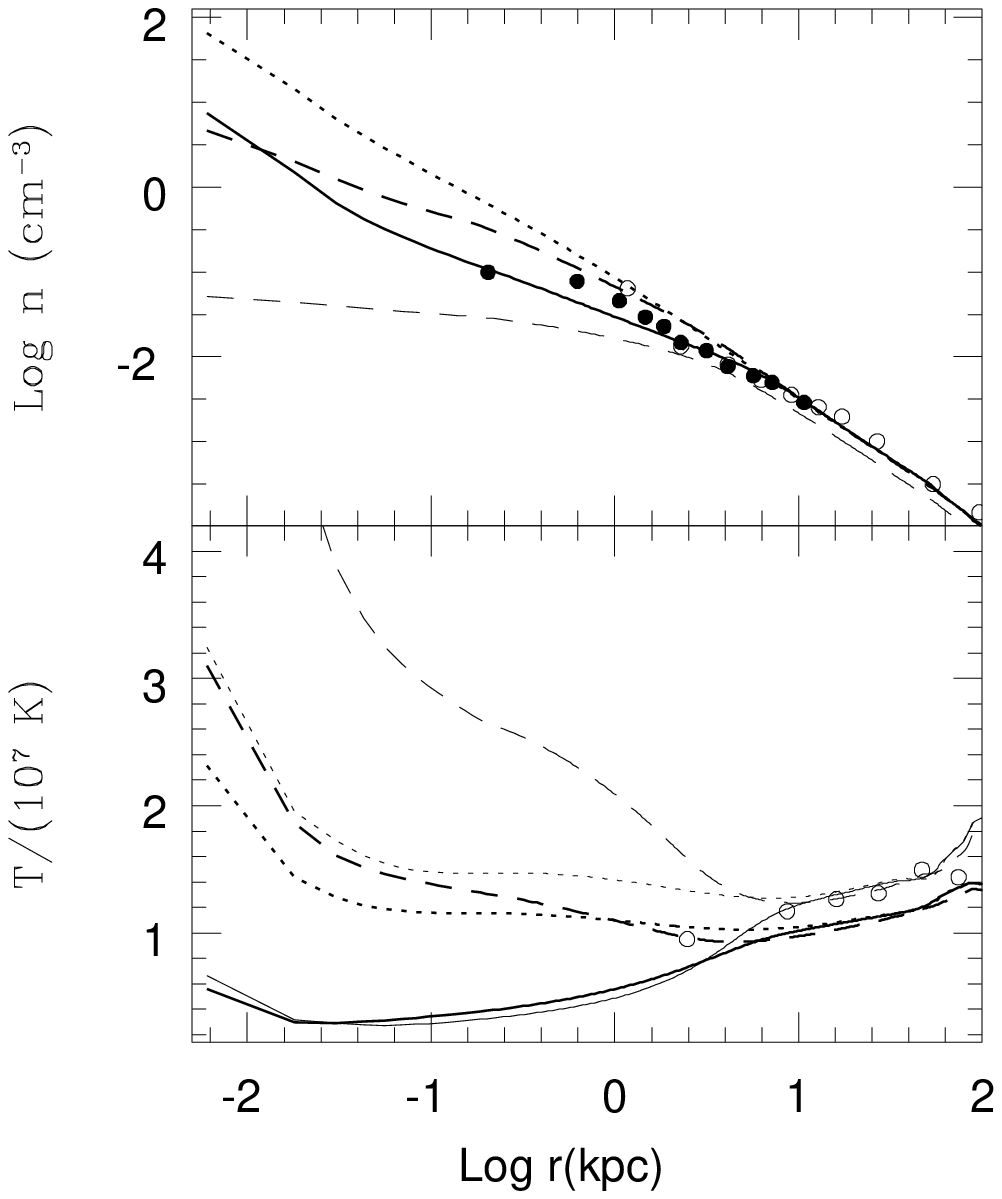]{
Three additional cooling flow solutions: with 
lower SMBH of mass $M_{bh} = 10^9$ $M_{\odot}$ 
({\it dotted lines}); with severe dropout 
$q(r) = \exp(n(r)/0.006)$ ({\it dashed lines});
and with extra pressure 
$P_{tot} = P(1 + 5\exp(-r/2{\rm kpc})$
({\it solid lines}).  
{\it Top Panel:} Radial variation of apparent gas density
including cooling regions; for the flow with severe dropout 
the background flow is also shown ({\it light dashed line}).
{\it Bottom Panel:} Radial variation of gas temperature 
in the background flow ({\it light lines}) 
and the combined temperature of the 
background flow and cooling regions 
viewed in projection ({\it heavy lines}).
Nomenclature for the observations is the same as 
in Fig. 1.
\label{fig3}}

\end{document}